\documentclass[11pt,twoside]{article}

\usepackage{asp2006}
\usepackage{adassconf}
\usepackage{epsfig}

\paperID{P64}
\paindex{Santander-Vela, J.D.}
\aindex{Delmotte, M.}
\aindex{Delgado, A.}
\aindex{Vuong, M.}


\begin{document}
	\title{Data Provenance: Use Cases for the ESO archive,
	and Interactions with the Virtual Observatory}
	\author{Juan de Dios Santander-Vela,
	Arancha Delgado,  Nausicaa Delmotte,
	and Myha Vuong}
	\affil{European Southern Observatory,
	Karl-Schwarzschild-Stra\ss{}e 2,
	Garching bei M\"unchen, D-85748, Germany}

\begin{abstract}
	In the Virtual Observatory era, where we intend to expose
	scientists (or software agents on their behalf) to a stream of
	observations from all existing facilities, the ability to
	access and to further interpret the origin, relationships, and
	processing steps on archived astronomical assets (their
	Provenance) is a requirement for proper observation selection,
	and quality assessment. In this article we present the different
	use cases Data Provenance is needed for, the challenges
	inherent to building such a system for the ESO archive,
	and their link with ongoing work in the International Virtual
	Observatory Alliance (IVOA).
\end{abstract}

\section{The Provenance of Astrophysical Data}
	The ever increasing datasets archived and dealt with by 
	astrophysical institutions and researchers are moving
	astrophysics into one of the fastest growing, and more public,
	e-Science disciplines.
	
\begin{figure}[tb]
		\centering
			\includegraphics[width=\textwidth]
			{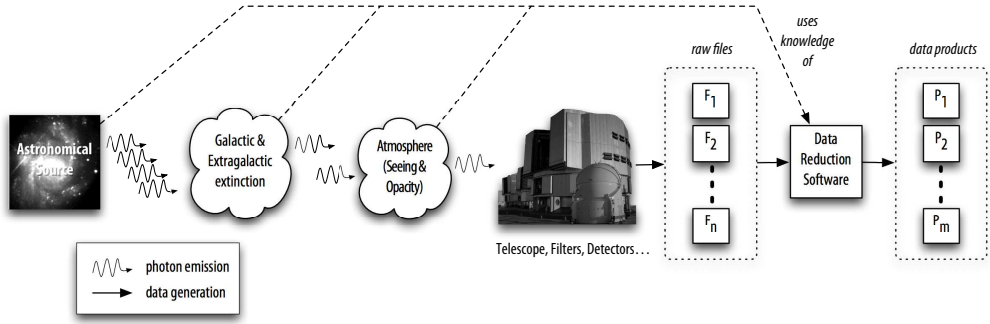}
		\caption{
			Data
			reduction software are
			convert properties measured with the telescope
			detectors into files ($F_n$), and later into
			physical parameters and data products ($P_m$).
			Knowledge and models of telescope, detectors
			settings, astronomical source, and
			absorbing media is needed for the reduction.
		}
		\label{fig:P64.santander.fig1}
	\end{figure}

	Within e-Science, Data Provenance is the tracking of the origin of
	any piece of information that has been recorded, with knowledge
	of all processing steps undergone. In the astronomical realm,
	we need to track data Provenance in order to answer questions like
	\emph{what was done to the photons captured by this archived
	entity}, and \emph{what entities were responsible for it?} 
	We need to be able to answer those questions because for typical
	astronomical workflows across large datasets Provenance must be
	recorded, and later used throughout the workflow, in order to get
	physically meaningful and accurate results.
	Fig.~\ref{fig:P64.santander.fig1} illustrates the role of
	Data Provenance in astronomical workflows.
	
	In particular, relevant questions for Data Provenance are
	\emph{which
	were the files used to create a particular data product}
	(Associations), and \emph{which are the data products created
	from it}? (Inverse Association).
	Additionally, we need to know
	\emph{are we using the same source data} (photons) \emph{for
	any two given data products}? (Photon Accountancy), as the use
	of the same photons in different data products which are later
	combined can produce unreliable science.
	Finally, we need to be
	able to \emph{track parameters related to the models for
	emission, absorption and detection} used by the reduction
	software, as these encode implicit knowledge that needs to be
	made explicit in order to allow the reproduction of
	measurements and processes by other scientists.
	
\section{Data Provenance Systems' Requirements:
High-Level Use Cases}
\label{sec:use_cases}
\begin{figure}[tb]
		\centering
			\includegraphics[width=\textwidth]
			{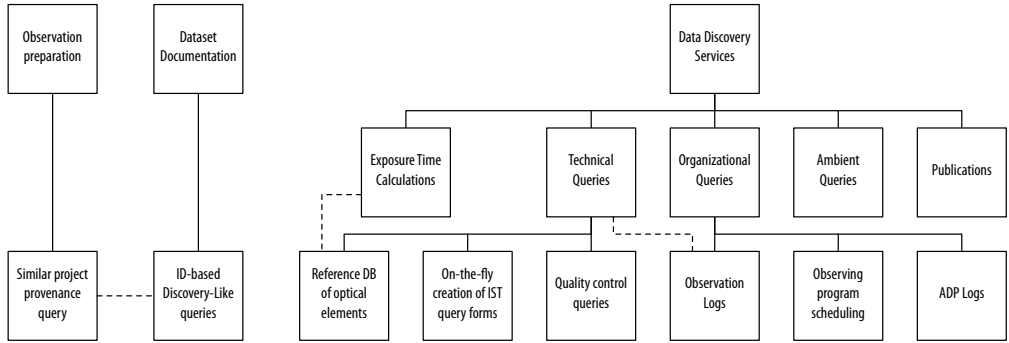}
		\caption{
			Classification of use cases compiled for the ESO archive.
			The main group corresponds to Data Discovery using
			Provenance metadata (mostly  through inverse
			association queries), while similar queries are
			desirable for dataset documentation (using direct
			association queries).
		}

		\label{fig:P64.santander.fig2}
	\end{figure}

	The use cases for a Data Provenance system within the ESO
	archive (see Fig.~\ref{fig:P64.santander.fig2}) have been
	classified in three kinds of services:
	
	\begin{description}
		\item[Data Discovery services] which query on the
		Provenance records to find scientific data fulfilling
		different criteria. These data discovery encompasses
		the Provenance of observing projects at all stages:
		Organization (observing proposals, PIs and COIs),
		Technical (instrument setup), Ambient, and Publications.
		In addition, Exposure Time Calculators (ETCs) must rely on
		the
		historical information of telescope elements, such as
		filters or detectors, whose properties (or actual
		elements) can change with time.
		
		\item[Dataset Documentation services] which allow scientists
		to retrieve the full record of observation origin and
		further processing. They are the inverse of the services
		above, as the dataset is already known, and the inverse
		association must be constructed. ETCs can be used in this
		family of use cases to assess data quality for a particular
		dataset.
		
		\item[Observation Preparation services] where both
		observations and projects with given data acquisition
		characteristics are retrieved. 
	\end{description}

\section{Implementing Data Provenance at ESO}
\subsection{Challenges}
	Following the taxonomy of Data Provenance systems by Simmhan,
	Plale, \& Gannon (2005), the main elements to be specified for
	Data Provenance systems are Storage, Granularity,
	Representation, and Dissemination.
	
	\begin{description}
		\item[Storage] deals with the actual way in which
		Data Provenance is made persistent. While Storage
		implementation is not relevant from an interoperability
		point of view, it provides the foundation for all
		Provenance services. Insertion and retrieval speeds
		must be able to cope with the expected data generation
		and query rate.

		\item[Granularity] is the degree of detail for which
		Provenance metadata is collected. It must be chosen
		to serve all ESO use cases, and also imposes
		restrictions on Storage (to be able to cope with the
		chosen Granularity). We can also include the selection
		of which particular metadata (from FITS headers to
		observation logs) are committed to Storage as part of
		the Granularity. From Fig.~\ref{fig:P64.santander.fig2}
		we can see that other metadata, such as Observation logs,
		and ADP processing logs, must be taken into account,
		and provide a time dimension to the problem.
		
		 \item[Representation] the way Provenance is expressed
		when shared with requesting entities (as opposed to
		Storage). Hierarchical trees, series of keyword/value
		pairs, or even plain-text files for informational use
		cases. However, due to the rich nature of the objects
		described by Provenance ---a detailed description of the
		light-path in ESO instruments is shown by Delgado
		(2009)---, a hierarchical structure is preferred.
		Representation directly affects interoperability, so
		the generalization of a Provenance representation as
		an IVOA Data Model would be desirable.
		
		 \item[Dissemination] the way Provenance metadata is
		to be available for shared datasets. Decisions in this
		respect include creating Provenance services to satisfy
		the use cases outlined in Section~\ref{sec:use_cases}
	 \end{description}
	
	The challenges, then, lie in successfully managing the huge
	datasets involved, do so maintaining system responsiveness, and
	standardize the way to retrieve and disseminate Provenance
	metadata.
	
\subsection{Jump-Starts}
	In order to find the best solution for some of the issues
	outlined above, we can count on several already started
	initiatives.
	
	 There are active discussions on the Provenance of
	astrophysical datasets within the IVOA Working Group
	for Data Modelling (DM WG). A Representation, in the form of a
	complete IVOA Observation data model, has already been proposed
	by Santander-Vela (2009), and is currently under study.
	
	 In addition, ESO maintains a complete FITS keyword/metadata
	database (Vuong et al. 2008), which will be one of the basis
	for establishing Storage and Granularity of the future ESO
	Provenance services. Currently, that facility stores more than
	five thousand million keyword-value pairs, for more than ten
	million raw file entries, and already includes metadata
	versioning support (a requirement for Provenance audit trails
	and historical dataset documentation), while at the same time
	providing excellent query speed. There is also an ongoing
	project for using that same database to ingest keywords for
	highly processed datasets.

\section{Conclusions}
	By identifying the elements needed for a successful data
	Provenance management system, and the selection of already in
	place systems, the ESO archive will be able to provide even
	more advanced scientific-oriented data products and services,
	while at the same time benefiting from existing IVOA standards
	for the Virtual Observatory, and helping in their development,
	leading by example.

\acknowledgements JDSV is grateful to the support 
	on his Ph.D. thesis by Spanish MICINN grants
	AYA2005-07516-C02-00 and AYA 2008-06181-C02-02; to the ESO
	Archive Department for support to this conference; and to Paul
	Eglitis for valuable comments. AD work is part of the Spanish
	in-kind contribution to ESO, and was supported by MICINN grant
	CAC-2006-47.

\end{document}